\documentclass[
  aps,
 prb,
  twoside,
  onecolumn,
  reprint,
  showpacs,
  floatfix,
  10pt,
  citeautoscript
]{revtex4-1}

\usepackage[utf8]{inputenc}
\usepackage[english]{babel}
\usepackage{graphicx}
\usepackage{multirow}
\usepackage{float}
\usepackage{units}
\usepackage{amsmath}
\usepackage{subfigure}
\usepackage[breaklinks]{hyperref}
\usepackage[hyphenbreaks]{breakurl}
\usepackage{wasysym}

\usepackage{dcolumn}
\newcolumntype{d}[1]{D{.}{.}{#1}}
\newcolumntype{e}[1]{D{-}{\text{-}}{#1}}
\newcolumntype{p}[1]{D{,}{\,\pm\,}{#1}}

\newcommand{\Eq}[1]{Eq.~(\ref{#1})}

\newcommand{\fig}[1]{Fig.~\ref{#1}}
\newcommand{\tab}[1]{Table~\ref{#1}}
\newcommand{\Sect}[1]{Section~\ref{#1}}

\newcommand{\BZO}{\text{BaZrO$_3$}}
\newcommand{\kb}{\text{$k$}}
\newcommand{\kbt}{\text{$kT$}}

\newcommand{\mue}{\text{$\mu_{\text{e}}$}}
\newcommand{\mugs}{\text{$\mu$}}
\newcommand{\Evbm}{\text{$\epsilon_{\text{VBM}}$}}
\newcommand{\Ecbm}{\text{$\epsilon_{\text{CBM}}$}}
\newcommand{\Egap}{E_{\text{gap}}}
\newcommand{\etal}{\textit{et al.}}
\newcommand{\V}{\text{v}}
\newcommand{\Ox}{\text{O}_\text{O}^{\times}}
\newcommand{\VOpp}{\text{v$_\text{O}^{\bullet\bullet}$}}
\newcommand{\OHp}{\text{OH$_\text{O}^{\bullet}$}}
\newcommand{\IP}{\text{IP}}
\newcommand{\EA}{\text{EA}}

\renewcommand{\O}{\text{O}}
\newcommand{\Ot}{\text{O$_2$}}
\renewcommand{\H}{\text{H}}
\newcommand{\HtO}{\text{H$_2$O}}
\newcommand{\h}{\text{h}}
\newcommand{\e}{\text{e}}
\newcommand{\Hox}{\text{$\Delta H_{\text{ox}}^{\circ}$}}
\newcommand{\Hoxeff}{\text{$\Delta H_{\text{ox}}^{\circ,\text{eff}}$}}
\newcommand{\Hhydr}{\text{$\Delta H_{\text{hydr}}^{\circ}$}}
\newcommand{\Sox}{\text{$\Delta S_{\text{ox}}^{\circ}$}}
\newcommand{\Shydr}{\text{$\Delta S_{\text{hydr}}^{\circ}$}}
\newcommand{\Shydreff}{\text{$\Delta S_{\text{hydr}}^{\circ,\text{eff}}$}}

\newcommand{\Eox}{\text{$\Delta E_{\text{ox}}$}}
\newcommand{\Ehydr}{\text{$\Delta E_{\text{hydr}}$}}
\newcommand{\Ec}{\varepsilon^{\text{coh}}}
\newcommand{\Ezp}{\varepsilon^{\text{Z.P.}}}

\newcommand{\Khydr}{\text{$K_{\text{hydr}}$}}
\newcommand{\Kox}{\text{$K_{\text{ox}}$}}
\newcommand{\GW}{\text{$G_0W_0$}}
\newcommand{\DEV}{\text{$\Delta E_{\text{v}}$}}
\newcommand{\DEH}{\text{$\Delta E_{\text{H}}$}}
\newcommand{\DUO}{\text{$\Delta U_{\text{O}}^{\text{vib}}$}}
\newcommand{\DUH}{\text{$\Delta U_{\text{H}}^{\text{vib}}$}}
\newcommand{\DSO}{\text{$\Delta S_{\text{O}}^{\text{vib}}$}}
\newcommand{\DSH}{\text{$\Delta S_{\text{H}}^{\text{vib}}$}}

\newcommand{\PBExGW}{\text{PBE$+\chi$[\GW]}}
\newcommand{\meff}{\text{$m_\text{h}^*$}}

\usepackage{color}

\usepackage{etoolbox}
\apptocmd{\sloppy}{\hbadness 10000\relax}{}{}

\begin{document}

\title{
  Implications of the band gap problem on
  oxidation  and hydration in acceptor-doped barium zirconate
}

\author{Anders Lindman}
\email{anders.lindman@chalmers.se}
\author{Paul Erhart}
\author{G\"oran Wahnstr\"om} 
\email{goran.wahnstrom@chalmers.se} 

\affiliation{
  Department of Applied Physics,
  Chalmers University of Technology,
  SE-412 96 Gothenburg, Sweden
}


\begin{abstract}

Charge carrier concentrations in acceptor-doped proton-conducting perovskites are to a large extent determined by the hydration and oxidation of oxygen vacancies, which introduce protons and holes, respectively. 
First-principles modeling of these reactions involves calculation of formation energies of charged defects, which requires an accurate description of the band gap and the position of the band edges.
Since density-functional theory (DFT) with local and semi-local exchange-correlation functionals (LDA and GGA) systematically fails to predict these quantities this can have serious implications on the modeling of defect reactions.
In this study we investigate how the description of band gap and band edge positions affects the hydration and oxidation in acceptor-doped \BZO.
First-principles calculations are performed in combination with thermodynamic modeling in order to obtain equilibrium charge carrier concentrations at different temperatures and partial pressures.
Three different methods have been considered: DFT with both semi-local (PBE) and hybrid (PBE0) exchange-correlation functionals, and many-body perturbation theory within the \GW-approximation. 
All three methods yield similar results for the hydration reaction, which are consistent with experimental findings.
For the oxidation reaction, on the other hand, there is a qualitative difference. 
PBE predicts the reaction to be exothermic while the two others predict an endothermic behavior.
Results from thermodynamic modeling are compared with available experimental data, such as enthalpies, concentrations and conductivities, and only the results obtained with PBE0 and \GW, with an endothermic oxidation behavior, give a satisfactory agreement with experiments. 

\end{abstract}

\maketitle
   
\section{Introduction}

Since the beginning of the 1980s, when Iwahara \etal \cite{iwahara_proton_1981} discovered proton conduction in acceptor-doped SrCeO$_3$, perovskite oxides ($AB$O$_3$) have been studied extensively with respect to their potential as proton conductors  \cite{kreuer_proton_2003,iwahara_prospect_2004,norby_proton_2009}. 
Such materials have many applications, including fuel cells, electrolyzers, hydrogen separation membranes and hydrogen sensors \cite{iwahara_prospect_2004}. 
Many acceptor-doped perovskites are also oxide ion, electron and hole conductors \cite{norby_solid-state_1999,phair_review_2006,sunarso_mixed_2008}
and suitable for applications such as electrodes and hydrogen separation membranes \cite{norby_solid-state_1999,phair_review_2006}.
As some applications rely on the perovskite being a pure ionic or electrical conductor while others do not it becomes important to understand and control the conductivity mechanisms in order to predict and optimize the material performance.

Proton incorporation into the perovskite structure is made possible through acceptor doping. By substituting $B$-site cations with dopant ions of lower valency positively charged oxygen vacancies are formed due to charge compensation. By exposing the doped perovskite to water vapor the vacancies can be filled by water molecules which introduces protons into the structure. 
In Kr{\"o}ger-Vink notation this reaction is expressed as
\begin{equation}\label{eq:hydration}  
\HtO(\text{g}) + \VOpp + \Ox \rightleftharpoons 2\OHp,
\end{equation}
which describes how a water molecule, an oxygen vacancy and an oxide ion form two hydroxide ions (protons).
Oxygen vacancies also enable the incorporation of holes, which can be introduced through oxidation of oxygen vacancies,
\begin{equation}\label{eq:oxidation}
\frac{1}{2}\Ot(\text{g}) + \VOpp \rightleftharpoons 2\h^{\bullet} + \Ox.
\end{equation}

Theoretical modeling based on density-functional theory (DFT) has become an important computational tool in materials science. 
The local density approximation (LDA) and various semi-local generalized gradient approximations (GGAs) are routinely being used.
In condensed matter research the Perdew-Burke-Ernzerhof \cite{perdew_generalized_1996,*perdew_generalized_1997} (PBE) type of GGA is currently the most common parametrization \cite{burke_perspective_2012}. 
GGAs have been applied to study hydration\cite{bjorketun_structure_2007,bevillon_hydration_2008,bjorheim_combined_2010,tauer_theoretical_2011,hermet_thermodynamics_2012,dawson_first-principles_2015,bjorheim_hydration_2015} as well as oxidation \cite{sundell_thermodynamics_2006,hermet_thermodynamics_2012,wang_atomistic_2009,bevillon_oxygen_2011} in different perovskite oxides.

For the oxidation process the semi-local GGAs predict the reaction to be exothermic \cite{sundell_thermodynamics_2006,hermet_thermodynamics_2012,wang_atomistic_2009,bjorheim_combined_2010,bevillon_oxygen_2011}: the hole concentration is decreasing with increasing temperature.
The hole conductivity is proportional to both the hole concentration and the hole mobility, and it is experimentally established that the hole conductivity increases with temperature \cite{bohn_electrical_2000,wang_ionic_2005,nomura_transport_2007,kuzmin_total_2009}. 
For the GGA result to be consistent with the experimental results it has therefore been suggested that the hole mobility increases more rapidly than the decrease in hole concentration \cite{hermet_thermodynamics_2012,bevillon_oxygen_2011}.
This, however, is not in line with the common view in the research field and it has been stated that the electronic structure of acceptor-doped proton-conducting perovskites remain surprisingly poorly understood \cite{kim_moving_2015}.

It is well-known that local and semi-local functionals underestimate the band gap of semiconductors and insulators \cite{perdew_density_1985,mori-sanchez_localization_2008};
a shortcoming that extends to the description of the valence and conduction band edges.
The position of the top of the valence band is decisive for a correct description of the oxidation reaction in \Eq{eq:oxidation} and hence one has to go beyond standard DFT with LDA/GGA in order to describe hole conduction properly. 

In this paper we have performed theoretical modeling of hydration and oxidation of an acceptor-doped oxide. 
Three defects are of interest in this context, the doubly positively charged oxygen vacancy, the hole and the proton, where the latter is often regarded as a hydroxide ion. 
Here we treat the hole as a band state. The methodology is applied to acceptor-doped \BZO, one of the most promising proton conducting perovskites since it combines high bulk proton conductivity with chemical stability \cite{kreuer_proton_2003,yamazaki_high_2009,fabbri_towards_2011}. 

First-principles calculations are used to determine the electronic structure and defect formation energies. 
The starting point is DFT based on the PBE functional for the exchange-correlation energy. 
To remedy the band gap problem we then consider two different approaches. 

The first one is based on a many-body perturbation technique. 
We determine the quasiparticle correction to the PBE energy levels using the $G_0W_0$-approximation introduced by Hedin \cite{hedin_new_1965,*HedLun70,*AulJoenWil99}. 
The second approach is based on hybrid functionals that admix a fraction of non-local exchange to a semi-local approximation.
We use the hybrid functional PBE0, which is obtained from PBE by replacing 25\%\ of the PBE exchange energy by Hartree-Fock exchange. 
To make our study less empirical we stick to this original suggestion of 25\%\ Hartree-Fock exchange \cite{perdew_rationale_1996} and we do not make use of range-separation, as introduced in the corresponding HSE functionals \cite{heyd_hybrid_2003,*heyd_erratum_2006}. 
Additionally, PBE0 has been shown to give a good description of \BZO \cite{evarestov_hybrid_2011}.

Thermodynamic modeling based on our first-principles results is then performed to obtain charge carrier concentrations in the acceptor-doped system at different temperatures and environmental conditions.
We find a qualitative difference for the oxidation reaction, being exothermic with PBE and endothermic using the \GW\ approach and the hybrid functional. 
Indeed, only the latter behavior is found to be consistent with experimental data of charge carrier concentrations and hole conductivities.

The paper is organized as follows. \Sect{sec:theory} describes the different aspects of the theoretical framework used in the paper while \Sect{sec:comp} contains the computational details of the PBE, PBE0 and \GW\ calculations. 
The results are presented and discussed in \Sect{sec:results} and \ref{sec:discussion}, and finally, a summary of the paper together with conclusions is given in \Sect{sec:summary}. 
The Appendix gives a description of band structure alignment with respect to the vacuum level based on surface calculations.

\section{Theory}\label{sec:theory}
In this work we study the thermodynamics of defect configurations in the dilute limit.
To this end, the formation free energies of individual point defects are calculated (if necessary for different charged states) as a function of atomic and electronic chemical potentials. 
The properties of the real system, most importantly defect concentrations, are then obtained by invoking the charge neutrality condition, which is employed to fix the electronic chemical potential under different environmental conditions (atomic chemical potentials).
An extensive review on the subject of first-principles modeling of defect formation in solids can be found in Ref.~\onlinecite{freysoldt_first-principles_2014}.

\subsection{Defect formation energies}

The formation energy of a defect in charge state $q$ is given by
\begin{align}
\Delta E_{\text{def}} &= E_{\text{def}}^{\text{tot}} + E_{\text{corr}}^q - E_{\text{id}}^{\text{tot}} - \sum_i\Delta n_i\mugs_i \nonumber \\
\label{eq:Eform}
&+ q(\Evbm + \mue + \Delta v^q),
\end{align}
where $E_{\text{def}}^{\text{tot}}$ and $E_{\text{id}}^{\text{tot}}$ are the total energies of the defective and ideal systems, respectively. $\Delta n_i$ denotes the change in atomic species $i$ upon defect formation and $\mugs_i$ is the corresponding chemical potential. Finally, $\mue$ represents the electron chemical potential with respect to the valence band maximum, $\Evbm$.
The terms $E_{\text{corr}}^q$ and $\Delta v^q$ are corrections that compensate errors associated with charged defects \cite{komsa_finite-size_2012}. 
The former term corrects errors due to image charge interactions, which are consequences of the periodic boundary conditions. 
The latter so-called potential alignment term corrects for the offset of electrostatic potentials of the charged defective and neutral ideal system.

The band gap problem of DFT affects the formation energies and can be approximately corrected for by using quasi-particle energy shifts from \GW\ calculations. 
The method considered here, which is a perturbative approach based on the DFT result, corresponds to applying a band gap correction to \Eq{eq:Eform} and is described in more detail in Refs.~\onlinecite{PerZhaLan05, AbeErhWil08,peng_convergence_2013}. 
In general, this approach requires knowledge of the shifts of both band edges as well as defect levels.
Fully ionized defects, which is the nature of the defects in this paper, are only affected by the shift of the valence band edge. 
The band gap corrected formation energy for such defects is given by
\begin{equation}
\Delta E_{\text{def}}^{\text{DFT}+\chi[GW]} = \Delta E_{\text{def}}^{\text{DFT}} + q\Delta\Evbm, 
\end{equation}          
where $\Delta\Evbm = \epsilon_{\text{VBM}}^{GW}-\epsilon_{\text{VBM}}^{\text{DFT}}$. 

For finite temperatures and pressures, \Eq{eq:Eform} can be written as    
\begin{align}
\Delta G_{\text{def}} &= G_{\text{def}}^{\text{tot}} + E_{\text{corr}}^q - G_{\text{id}}^{\text{tot}} - \sum_i\Delta n_i g_i \nonumber \\
\label{eq:Gform}
&+ q(\Evbm + \mue + \Delta v^q),
\end{align}
where $G_{\text{def}}^{\text{tot}}$ and $G_{\text{id}}^{\text{tot}}$ are the Gibbs free energies of the defective and ideal systems respectively, and $g_i$ is the chemical potential of the elemental reference phase $i$ at finite temperatures and pressures. 

\subsection{Chemical potentials of the gas phase}\label{sec:chempots}

The considered defects are oxygen vacancies and protons and chemical potentials for O and H are therefore needed.
The environments of the oxidation and hydration reaction are oxygen gas (\Ot) and water vapor (\HtO), and the chemical potentials of O and H are thereby expressed as
\begin{align}
g_{\O} &= \frac{1}{2}g_{\O_2} \\
g_{\H} &= \frac{1}{2}g_{\HtO}-\frac{1}{4}g_{\Ot}.
\end{align}
By assuming an ideal gas behavior the chemical potential of \Ot\ at temperature $T$ and partial pressure $p_\Ot$ (and equivalently for \HtO\ at $p_\HtO$) can be written as 
\begin{align}
g_{\Ot}(T,p_\Ot) &= \mugs_\Ot + \Ezp_{\Ot} +  h_\Ot^{\circ}(T) - Ts_\Ot^{\circ}(T) \nonumber \\ \label{eq:chempot}
&+ \kbt\ln{\frac{p_\Ot}{p_\Ot^{\circ}}}, 
\end{align}
where $\Ezp_\Ot$ is the zero-point energy of the \Ot\ molecule and $h_\Ot^{\circ}(T)$ and $s_\Ot^{\circ}(T)$ represent the temperature dependencies of enthalpy and entropy of the gas phase at the reference pressure $p_\Ot^{\circ}$. 
The enthalpies and entropies of \Ot\ and \HtO\ are extracted from thermodynamic tables \cite{chase_nist_1985}. 
Within the harmonic approximation the zero-point energies are given by $\sum_k\hbar\omega_k/2$, where $\omega_k$ are the molecular vibrational frequencies. 
Experimentally determined frequencies \cite{herzberg_molecular_1979,shimanouchi_tables_1977} yield $\Ezp_\Ot = \unit[0.10]{eV}$ and $\Ezp_\HtO = \unit[0.56]{eV}$.

Total energies from DFT are used for $\mugs_i$. Common practice is to use the molecular total energies
\begin{align}
\mugs_{\Ot} &= E_{\Ot}^{\text{tot}} \\
\mugs_{\HtO} &= E_{\HtO}^{\text{tot}}.
\end{align}
This is problematic since PBE is known to overbind the \Ot\ molecule with \unit[0.9]{eV}. 
To overcome this problem total energies of atoms are used instead and combined with experimental values for the cohesive energies $\Ec$ according to
\begin{align}
\mugs_{\Ot} &= 2E_{\O}^{\text{tot}} + \Ec_{\Ot} \\
\mugs_{\HtO} &= 2E_{\H}^{\text{tot}} + E_{\O}^{\text{tot}} + \Ec_{\HtO} .
\end{align}
With experimental data from Ref.~\onlinecite{chase_nist_1985} we obtain $\Ec_{\Ot}=\unit[-5.21]{eV}$ and $\Ec_{\HtO}=\unit[-10.07]{eV}$, where the zero-point energies (see above) have been removed. 

\subsection{Free energy of the solid phase}

The considered expression for the free energy of the solid phase depends only on temperature since the $PV$-term is very small within this context and can be neglected.
This implies that the Gibbs and Helmholtz free energies are practically identical and one can write
\begin{equation}
G(T) \approx F(T) = E^{\text{tot}} + U^{\text{vib}}(T) - TS^{\text{vib}}(T),
\end{equation} 
where $E^{\text{tot}}$ is the electronic contribution, and the temperature dependent terms $U^{\text{vib}}(T)$ and $S^{\text{vib}}(T)$ represent vibrational contributions. 
The latter two are calculated within the harmonic approximation using an Einstein model \cite{sundell_thermodynamics_2006,bjorketun_structure_2007}. 
Here we assume that the formation of a defect does not affect the vibrational frequencies of neighboring atoms.
The change in $U^{\text{vib}}(T)$ and $S^{\text{vib}}(T)$ due to the addition of one atom of species $i$ is given by
\begin{align}
\Delta U_i^{\text{vib}}(T) &= \sum_{k=1}^{3}\left( \frac{\hbar\omega_{i,k}}{2} + \frac{\hbar\omega_{i,k}}{e^{\hbar\omega_{i,k}/\kbt}-1} \right) \\
\Delta S_i^{\text{vib}}(T) &= \kb\sum_{k=1}^{3}\left[ \frac{\hbar\omega_{i,k}/\kbt}{e^{\hbar\omega_{i,k}/\kbt}-1} -\ln{\left(1-e^{-\hbar\omega_{i,k}/\kbt}\right)} \right],
\end{align}
where $\omega_{i,k}$ are the vibrational frequencies. 
For the oxygen atom we use the frequencies $\unit[557]{cm^{-1}}$, $\unit[250]{cm^{-1}}$ and $\unit[250]{cm^{-1}}$, and for the proton we use $\unit[3502]{cm^{-1}}$, $\unit[900]{cm^{-1}}$ and $\unit[601]{cm^{-1}}$, which have been extracted from Ref.~\onlinecite{sundell_thermodynamics_2006,bjorketun_structure_2007}. 

\subsection{Defect concentration}\label{sec:defconc}

Defect concentrations are considered to be within the dilute limit and are therefore given by 
\begin{equation}\label{eq:conc}
c_{\text{def}} = \frac{N_{\text{def}}}{V_c}e^{-\Delta G_{\text{def}}/\kbt},
\end{equation}
where $N_{\text{def}}$ is the number of defect sites in the primitive cell with volume $V_c$.
In this case $V_c=a_0^3$ with $a_0$ being the lattice constant. 
There are three oxygen sites in the primitive cell and therefore three available sites for the oxygen vacancy, i.e.,  $N_\V = 3$.
Proton sites are associated with oxygen ions, with four possible configurations per oxygen site \cite{bjorketun_structure_2007}, which yields $N_\H=12$ proton sites in each primitive cell. 

In order for the dilute-limit approximation to be valid the occupancy has to be much smaller than the number of available sites ($c_\text{def}V_c\ll N_\text{def}$). 
In this paper we use a dopant concentration of 10\%, which yields a maximum proton occupancy of 0.1 per primitive cell.  
This corresponds to 1 in 120 proton sites being occupied. 
The same dopant concentration yields an upper limit of 0.05 oxygen vacancies per primitive cell, which corresponds to 1 in 60 oxygen sites being vacant. 
The dilute-limit approximation is thus justified.

\subsection{Electron chemical potential}\label{sec:calcfermi}

The electron chemical potential \mue\ is obtained by solving the charge neutrality condition 
\begin{equation}
\sum_{\text{def}}qc_{\text{def}}(\mue) - n_\e(\mue) + n_\h(\mue) = 0,
\label{eq:charge_neutrality}
\end{equation}
where $n_\e$ and $n_\h$ are the electron and hole concentration, respectively, and the sum is over all defects in the material including the acceptor dopants.
Equation~(\ref{eq:charge_neutrality}) can be solved by iteration \cite{erhart_modeling_2008}. In the present work electrons and holes are treated as band states and the corresponding concentrations are obtained from the density of states (DOS) $g(\epsilon)$ according to
\begin{align}
n_{\e}  &= \int_{\Ecbm}^{\infty}g(\epsilon)f(\epsilon,\mue)d\epsilon \\
n_{\h}  &= \int_{-\infty}^{\Evbm}g(\epsilon)\left[1-f(\epsilon,\mue)\right]d\epsilon ,
\end{align}
where \Evbm\ and \Ecbm\ denote the positions of the valence band maximum (VBM) and conduction band minimum (CBM), respectively, and $f(\epsilon,\mue) = \{\exp{[(\epsilon-\Evbm-\mue)/\kbt]}+1\}^{-1}$ is the Fermi-Dirac distribution function.
The DOS is determined from first-principles calculations.

\section{Computational details}\label{sec:comp}

First-principles calculations within the density-functional theory (DFT) formalism were carried out using the Vienna \textit{ab-initio} simulation package \cite{kresse1993,*kresse1994,*kresse1996a,*kresse1996b}, which uses plane wave basis sets and periodic boundary conditions. 
The projector augmented wave method (PAW) \cite{bloechl1994,*kresse1999} was employed to describe ion-electron interactions. 
Two different functionals were used to model exchange and correlation in their non-spin polarized versions: the generalized gradient approximation functional PBE \cite{perdew_generalized_1996,*perdew_generalized_1997} and the hybrid functional PBE0 \cite{perdew_rationale_1996}.  
The plane wave cutoff energy was set to \unit[400]{eV} and a $6\times6\times6$ Monkhorst-Pack grid was used for $k$-point sampling of the \BZO\ primitive cell and then reduced accordingly with increasing supercell size. 
Super cells comprising up to $6\times6\times6$ unit cells were used for defect calculations based on the PBE functional. PBE0 calculations were conducted for $3\times3\times3$ supercells only. Ionic relaxation was carried for all structures until the residual forces were below \unit[0.02]{eV\AA$^{-1}$}. 

All calculations were performed with the cubic perovskite structure, which belongs to space group Pm$\bar{3}$m. 
The optimized PBE lattice constant of \unit[4.235]{\AA} is somewhat larger than experimental values \unit[4.191-4.197]{\AA}\cite{pagnier_neutron_2000,levin_phase_2003}, but in agreement with previous theoretical studies of \BZO\ based on GGA functionals \cite{sundell_thermodynamics_2006,bjorketun_structure_2007,bjorheim_combined_2010}. 
The PBE0 calculations were carried out at the PBE lattice constant for a more direct comparison.
 
Many-body calculations were carried out within the formalism of the quasi-particle method $GW$ \cite{hedin_new_1965,*HedLun70,*AulJoenWil99}. More specifically, the \GW\ approach was used.
Calculations were based on PBE wave functions and employed PAW data sets optimized for $GW$ calculations \cite{ShiKre06, *ShiKre07}. 
The general plane wave cutoff energy was 434\,eV while a cutoff of 290\,eV was employed in the response function calculations. 
The Brillouin zone was sampled using a $\Gamma$-centered $5\times5\times5$ $k$-point mesh and all calculations were carried out at the PBE lattice constant.

\begin{figure}
\begin{center}
\includegraphics[width=0.48\textwidth]{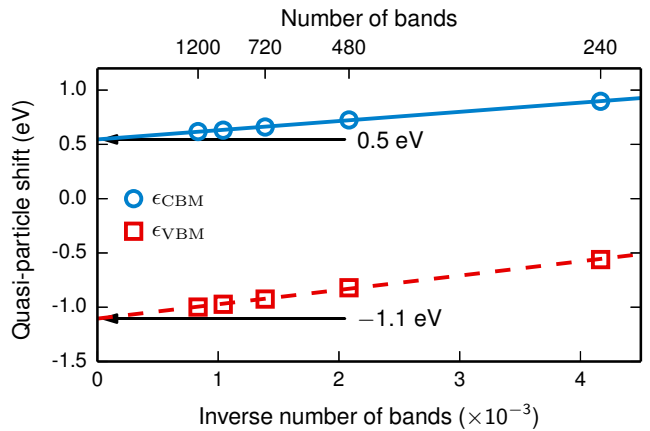}
\caption{Convergence of quasi-particle energies from \GW\ calculations based on PBE wave functions with respect to the number of bands included in the calculation.
}
\label{fig:GW_convergence}
\end{center}
\end{figure}

While the band gap converges relatively quickly with the number of empty states included in the calculations, individual quasi-particle energies typically converge more slowly. 
As shown in \fig{fig:GW_convergence} VBM and CBM are, however, observed to depend linearly on the inverse number of bands, whence converged values were obtained by extrapolation. 
This approach is similar to the hyperbolic fit employed in Ref.~\onlinecite{FriMulBlu11}.

\section{Results}\label{sec:results}
\subsection{Electronic structure}\label{sec:elstruct}

\begin{figure}
  \centering
  \includegraphics[width=0.48\textwidth]{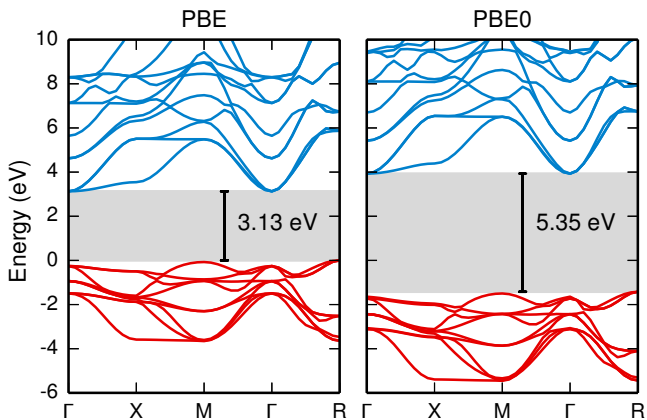}
  \caption{
    Comparison of PBE and PBE0 band structures for $\BZO$. Blue and red lines represent empty and occupied bands, respectively. The grey areas indicate the extent of the indirect band gap (R-$\Gamma$). The energy scale is chosen to be zero at the PBE VBM.
  }
  \label{fig:bs_ideal_comp}
\end{figure}

\begin{figure}
  \centering
  \includegraphics[width=0.48\textwidth]{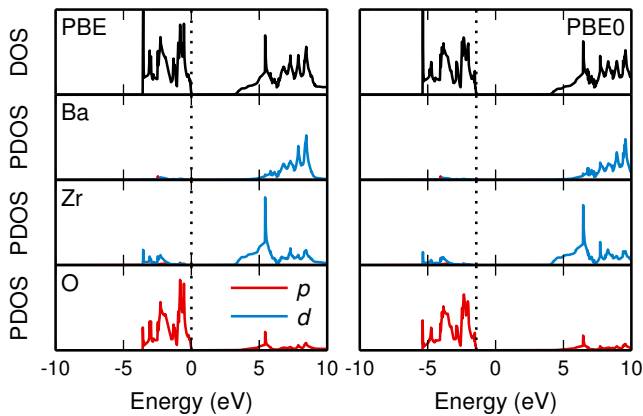}
  \caption{
    Total and partial density of states for $\BZO$ obtained with PBE and PBE0. The red and blue lines correspond to electronic $p$- and $d$-states and the dashed vertical line marks the VBM.
  }
  \label{fig:pdos_comp}
\end{figure}

The band structure of \BZO\ from PBE and PBE0 calculations is presented in \fig{fig:bs_ideal_comp}. 
The band gap is indirect with the VBM at R and the CBM at $\Gamma$. 
The size of the gap, which is determined from single-particle eigenvalues\cite{alkauskas_bandedge_2011}, is \unit[3.13]{eV} and \unit[5.35]{eV} with PBE and PBE0, respectively. 
The direct band gap, with the VBM and CBM at $\Gamma$, is only slightly larger: \unit[3.38]{eV} with PBE and \unit[5.57]{eV} with PBE0.
The shape of the band structures is very similar in both cases, which indicates that main difference between PBE and PBE0 lies in the size of band gap and the position of band edges. 
This is illustrated in \fig{fig:pdos_comp}, which shows the total (DOS) and partial density of states (PDOS).
The valence band consists of oxygen $p$-states while the conduction band of zirconium $d$-states.

To compare the position of band edges the PBE and PBE0 band structures need to be properly aligned. 
Such an alignment can be done with respect to common reference potential, e.g., the average local electrostatic potential or the vacuum level\cite{alkauskas_defect_2008,*alkauskas_defect_2011}. 
In this study we use the same pseudopotentials and lattice constant for both PBE and PBE0 and the ionic contribution to the electrostatic potential is therefore the same. 
The electron density is found to be very similar with both methods, which yield similar contributions to the potential as well. 
As a consequence the average local electrostatic potential is approximately the same for both methods and the two band structures should be properly aligned \cite{alkauskas_defect_2008,*alkauskas_defect_2011}. 
We have also performed alignment with respect to the vacuum level using surface calculations, which verifies this alignment (see Appendix).

\begin{figure}
\begin{center}
\includegraphics[width=0.48\textwidth]{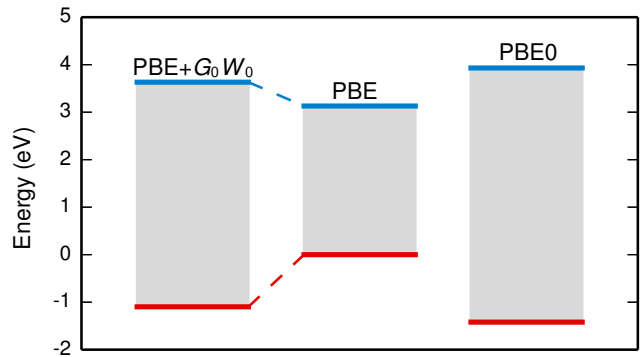}
\caption{Size and relative position of band gap for PBE, PBE0 and PBE+\GW\ calculations, where zero is set at the PBE VBM. The dashed lines indicate that PBE+\GW\ is a perturbative approach based on PBE.
Also see \tab{tab:bandgapcomp}.
}
\label{fig:bandgap}
\end{center}
\end{figure}

Band gaps and band edge positions are summarized in \tab{tab:bandgapcomp} and visualized in \fig{fig:bandgap}, where the band edge positions are given with respect to the PBE VBM. 
Both PBE+\GW\ and PBE0 open up the band gap, from \unit[3.13]{eV} to \unit[4.73]{eV} and \unit[5.35]{eV}, respectively, and yield VBM/CBM shifts that are qualitatively similar. 
The rather good agreement between PBE0 and PBE+\GW\ calculations for the VBM offset is not trivial as it has been shown that PBE0 band edge positions can differ quite substantially from \GW\ calculations, especially for wide band gap materials \cite{ChePas12}. In general one should expect PBE+\GW\ calculations to be more reliable for this purpose as they represent a more rigorous theoretical approach.

\begin{table}
\centering
\caption{Comparison of theoretical and experimental band gaps $\Egap$, as well as VBM and CBM shifts $\Delta\epsilon$ obtained from PBE0 and PBE+\GW\ calculations with respect to PBE calculations. All values are given in units of eV. The theoretical data are also visualized in \fig{fig:bandgap}.}
\begin{ruledtabular}
\begin{tabular}{l d{2.2} d{1.2} c}
Method	& \multicolumn{1}{c}{$\Delta\Evbm$} 	&  \multicolumn{1}{c}{$\Delta\Ecbm$}  	& $\Egap$ 	\\ \hline
PBE		& 				&				& $3.13$		\\
PBE+\GW		& -1.10			& 0.50			& $4.73$		\\ 
PBE0	& -1.42			& 0.80			& $5.35$		\\	
Experiment		&				& 				& $5.3$\cite{robertson_band_2000}, $4.86$\cite{cavalcante_experimental_2008,*cavalcante_intense_2008,*cavalcante_intense_2009}, $4.8$\cite{yuan_synthesis_2008}
\end{tabular}
\end{ruledtabular}
\label{tab:bandgapcomp}
\end{table}

There are several experimental values for the band gap of \BZO\ in the literature. 
Robertson\cite{robertson_band_2000} reports a value of \unit[5.3]{eV}, which in close agreement with the PBE0 result. 
More recent studies by Cavalcante \etal\ \cite{cavalcante_experimental_2008,*cavalcante_intense_2008,*cavalcante_intense_2009} and Yuan \etal\ \cite{yuan_synthesis_2008} report band gaps in the range \unit[4.8-4.9]{eV}, which agree better with PBE+\GW.
The fact that the PBE+\GW\ still slightly underestimates the experimental band gap is consistent with calculations on other wide band gap materials \cite{ShiKre07,ChePas12,AbeSadErh12}.

\subsection{Defect formation energies}\label{sec:results_defect}
Formation energies have been calculated for the oxygen vacancy \DEV\ and the proton \DEH. 
The considered charge state of the vacancy is $+2$, which is the relevant state for the oxidation and hydration reactions.

The terms $E_{\text{corr}}^q$ and $\Delta v^q$ in the expression for the formation energy (see \Eq{eq:Eform}) are corrections to errors introduced by charged defects and periodic boundary conditions. 
Several correction schemes have been proposed over the years to reduce these errors (see Refs.~\onlinecite{komsa_finite-size_2012,kumagai_electrostatics-based_2014} for examples). 
Here we employ the finite-size scaling approach, in which the formation energy is calculated for several different supercell sizes and the corrected value $E_\infty$ is obtained by fitting the data points to a polynomial of the form
\begin{equation}\label{eq:extrapolation}
E(N) = aN^{-1} + bN^{-1/3}+E_\infty,
\end{equation}
where $N$ is the number of atoms in the supercell. In this fashion not only the leading terms of the multipole expansion of the electrostatic image interaction \cite{MakPay95} are accounted for but also elastic image interactions \cite{GreDed71, *DedPol72}.
This approach is suitable in this case since it is computationally feasible to obtain a sufficiently large number of data points for a reasonable fit. 
Additionally, since the screening in \BZO\ is quite large (the static dielectric constant $\varepsilon_\text{r}$ has been experimentally measured to fall in the range 40--160\cite{chen_origin_2011}), electrostatic image charge interactions, which are proportional to $\varepsilon_\text{r}^{-1}$, can be expected to be small.
There is thus no benefit in using more advanced schemes.

\begin{figure}
\begin{center}
\includegraphics[width=0.48\textwidth]{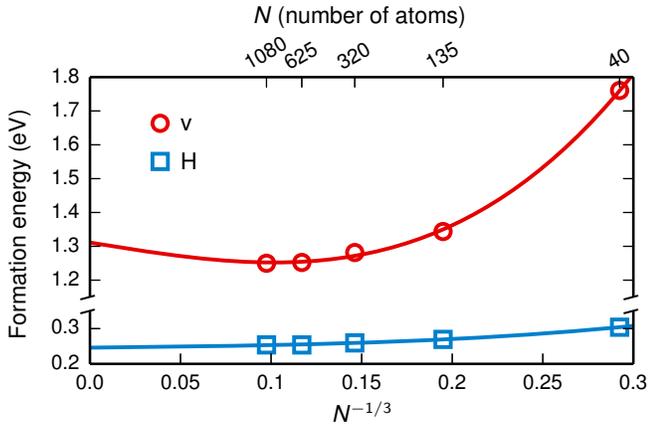}
\caption{Convergence of formation energies for the $+2$ charged oxygen vacancy (v) and the proton (H) with respect to supercell size (number of atoms). Solid lines are fits of the data to \Eq{eq:extrapolation}. The calculations are performed with the PBE functional.}
\label{fig:eformconv}
\end{center}
\end{figure}

Supercells with up to $6\times6\times6$ unit cells are considered for the extrapolation, which corresponds to 1080 atoms in the non-defective configuration. 
The results for the PBE functional are shown in \fig{fig:eformconv}. 
The extrapolated formation energy for the oxygen vacancy is $\unit[1.31]{eV}$ while a value of $\unit[0.25]{eV}$ is obtained for the proton. 
The figure shows that the formation energy of both defects is quite close to the extrapolated value already for $3\times3\times3$ supercells ($135\pm 1$ atoms), which is related to the strong electrostatic screening. 
Since the PBE formation energies of the $3\times3\times3$ supercell are already very close to the extrapolated value, this supercell size was employed for PBE0 calculations, which are computationally much more demanding.

Defect formation energies obtained from PBE and PBE0 calculations are summarized in \tab{tab:eformcomp}. 
All values are determined at the VBM corresponding to $\mue=0$. 
The differences between the PBE0 and PBE values are $\unit[-2.84]{eV}$ for the vacancy and $\unit[-1.47]{eV}$ for the proton. 
These differences are very close to $2\Delta\Evbm$ and $\Delta\Evbm$ (see \tab{tab:bandgapcomp}), which indicates that the difference between PBE and PBE0 is mostly due to the shift of the VBM. 
This observation in turn validates the \PBExGW\ approach. 

\begin{table}
\begin{center}
\caption{Comparison of formation (\DEV\ and \DEH) and reaction (\Eox\ and \Ehydr) energies, where the former are given for the electron chemical potential being located at the VBM ($\mue=0$). For PBE and \PBExGW\ the formation energies are extrapolated values, see \Eq{eq:extrapolation} and \fig{fig:eformconv}. PBE0 values correspond to $3\times3\times3$ supercells. All energies are given in units of eV.}
\begin{ruledtabular}
\begin{tabular}{l d{2.2} d{2.2} d{2.2} d{2.2}}
Method		& \multicolumn{1}{c}{\DEV}	& \multicolumn{1}{c}{\DEH}	& \multicolumn{1}{c}{\Eox}	& \multicolumn{1}{c}{\Ehydr}	\\ \hline
PBE			& 1.31					& 0.25					& -1.31					& -0.82					\\
\PBExGW		& -0.88					& -0.85					& 0.88				         & -0.82					\\
PBE0		& -1.53					& -1.22					& 1.53					& -0.90					\\
\end{tabular}
\end{ruledtabular}
\label{tab:eformcomp}
\end{center}
\end{table}

\subsection{Reaction enthalpies and entropies}\label{sec:reaction_ent}

The energy of the oxidation reaction in \Eq{eq:oxidation} is determined according to
 \begin{equation}\label{eq:oxenergy}
\Eox = 2\mue-\DEV(\mue),
\end{equation}
which is independent of \mue. 
Calculated values for \Eox\ are listed in \tab{tab:eformcomp}. 
With PBE the oxidation energy is $\unit[-1.31]{eV}$, which implies an exothermic reaction favoring the formation of holes. 
With \PBExGW\ and PBE0 the oxidation energy is $\unit[0.88]{eV}$ and $\unit[1.53]{eV}$, respectively, which corresponds to an endothermic reaction favoring oxygen vacancy formation.

The energy of the hydration reaction in \Eq{eq:hydration} is given by
 \begin{equation}\label{eq:hydrenergy}
\Ehydr = 2\DEH(\mue) - \DEV(\mue),
\end{equation}
which, like the oxidation energy, is independent of \mue. 
All three methods predict the reaction to be exothermic with a similar magnitude for \Ehydr, see \tab{tab:eformcomp}.
The reaction is slightly more energetically favorable with PBE0 compared to PBE, while PBE and \PBExGW\ yield identical values by construction.
This close agreement between the different methods can be traced to the fact that the hydration energy does not depend on the position of the VBM.

The standard enthalpy for both reactions can be determined from \Eox\ and \Ehydr\ by including the zero-point energies and the temperature dependence of both the solid and the gas phase. 
The enthalpies are given by
\begin{align}\label{eq:oxenthalpy}
\Hox(T) &= \Eox + \DUO(T) -  \frac{1}{2}\Ezp_\Ot  -  \frac{1}{2}h_\Ot^{\circ}(T) \\
\Hhydr(T) &= \Ehydr + 2\DUH(T) + \DUO(T) \nonumber \\
&- \Ezp_\HtO -  h_\HtO^{\circ}(T).
\end{align}
Similarly, the entropies are given by
\begin{align}
\Sox(T) &= \DSO(T) - \frac{1}{2}s_\Ot^{\circ}(T) \\
\Shydr(T) &= 2\DSH(T) + \DSO(T) -  s_\HtO^{\circ}(T).
\end{align}

\begin{figure}
\begin{center}
\includegraphics[width=0.48\textwidth]{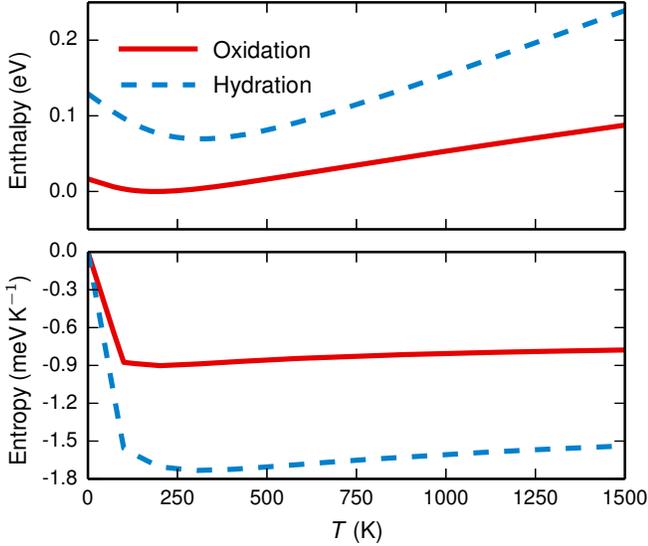}
\caption{Temperature dependence of the standard enthalpy and entropy of the hydration and oxidation reactions, see Equations (\ref{eq:hydration}) and (\ref{eq:oxidation}). The electronic contributions to the enthalpy ($\Eox$ and $\Ehydr$) have been subtracted, thus the values at zero temperature correspond to zero point energies. }
\label{fig:enttemp}
\end{center}
\end{figure}

In \fig{fig:enttemp} we show the standard enthalpy and entropy as a function of temperature for both reactions. 
\Eox\ and \Ehydr\ have been subtracted from the enthalpy thus the values at zero temperature correspond to the net zero-point energy of the reactions. 
These values are much less than the zero-point energy of the respective phases, which indicate that there is a large cancellation effect. 
Thus, if zero-point motion effects are included it is of importance to consider contributions from both the gas and solid phases. 

\subsection{Oxidation}

Based on the computed formation energies the equilibrium defect concentrations can be determined for different temperatures and pressures using the self-consistent scheme described in \Sect{sec:theory}. 
With these concentrations the oxidation reaction can be studied by calculating the corresponding equilibrium constant
\begin{equation}\label{eq:kox1}
\Kox(T) = \left( \frac{p_\Ot}{p_\Ot^{\circ}}\right)^{-1/2}\frac{n_\h^2c_\O}{c_\V},
\end{equation}
where $c_\V$ and $c_\O$ denote oxygen vacancy and oxygen ion concentrations, respectively. 

\begin{figure}
\begin{center}
\includegraphics[width=0.48\textwidth]{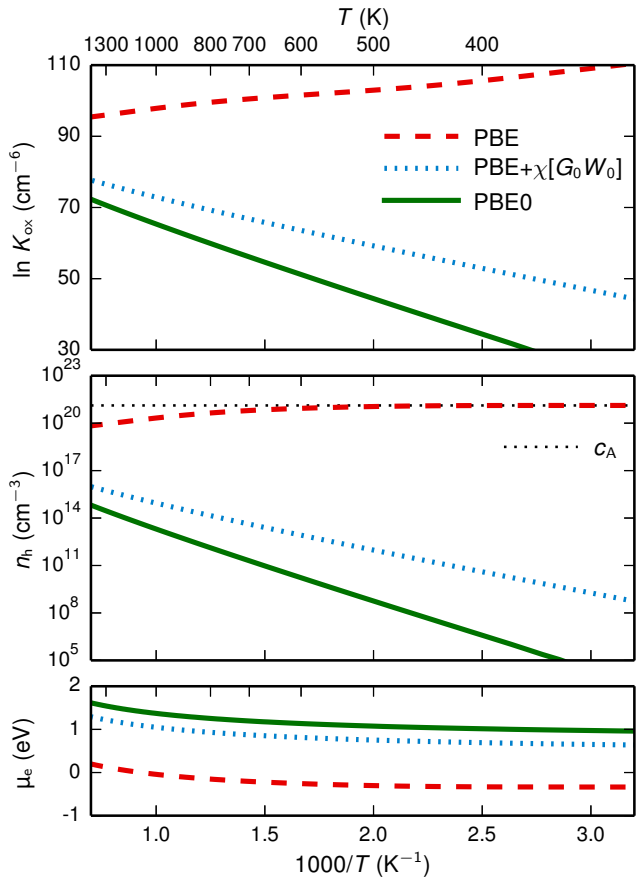}
\caption{The upper panel shows the equilibrium constant of the oxidation reaction in \Eq{eq:oxidation} while the middle and lower panels show the corresponding hole concentration and self-consistently obtained electron chemical potential. The concentrations are obtained with $p_{\Ot}=\unit[1]{bar}$ and a dopant concentration of 10\%, where the latter is depicted as a black dashed line in the middle panel.}
\label{fig:kox}
\end{center}
\end{figure}

In \fig{fig:kox} we show the equilibrium constant as function of temperature together with the hole concentration of a 10\%\ acceptor-doped system at the reference pressure ($p_{\Ot}=\unit[1]{bar}$). 
As can be expected from the oxidation enthalpies, the results differ quite significantly between PBE and the other two methods. 
With PBE the hole concentration increases with decreasing temperature and is completely compensating the dopant charge at lower temperatures. 
With PBE0 and \PBExGW\ the concentration displays the inverse temperature dependence and is several orders of magnitude smaller.
These features are reflected in the equilibrium constant, where the positive slope of the PBE curve indicates an exothermic process while the negative slope obtained using the other two methods corresponds to an endothermic reaction. 

In general, the slope of the $\ln{K(T)}$-curve is considered to correspond to the enthalpy of the reaction.
We define an effective oxidation enthalpy according to
\begin{equation}
\Hoxeff(T) = -\kb\frac{d\ln{\Kox(T)}}{d(1/T)}.
\label{eq:Hoexff}
\end{equation} 
Fitting the data in \fig{fig:kox} to \Eq{eq:Hoexff} yields $\Hoxeff(T=1000\,\text{K})$ values of $\unit[-0.66]{eV}$, $\unit[1.30]{eV}$ and $\unit[1.92]{eV}$ for PBE, \PBExGW\ and PBE0, respectively.
These values can be compared with $\unit[-1.26]{eV}$, $\unit[0.93]{eV}$ and $\unit[1.58]{eV}$ for $\Hox(T)$ at $T=\unit[1000]{K}$.

The electron chemical potential, which is also depicted in \fig{fig:kox}, is negative with PBE below \unit[1000]{K} and remains close to the valence band edge for larger temperatures. 
For PBE0 and \PBExGW\ on the other hand the electron chemical potential is located well within the band gap over the entire temperature range. 
In the latter case the Boltzmann approximation can be used to find a more simplified expression for $\Kox(T)$ and $n_\H$. 
The equilibrium constant can then be written as \cite{supplementary}
\begin{equation}
\Kox(T) =\left[n_\text{VB}(T)\right]^2e^{-\Hox(T)/\kbt}e^{\Sox(T)/\kb},
\end{equation}
where $n_\text{VB}(T)=2(\meff\kbt/2\pi\hbar^2)^{3/2}$ and $\meff$ is the effective mass for the hole. 
From this expression it follows \cite{supplementary} that 
\begin{equation}\label{eq:dlnkox}
\Hoxeff(T) = \Hox(T) + 3\kbt.
\end{equation} 
The contribution $3\kbt$ stems from the holes and is equal to \unit[0.26]{eV} at \unit[1000]{K}.
This explains the difference between the slopes of the \PBExGW\ and PBE0 curves in \fig{fig:kox} and the corresponding oxidation enthalpies $\Hox(T)$.
While for the PBE there is also a positive contribution to $\Hox(T)$ it is more difficult to obtain an explicit expression \cite{supplementary}.

\begin{figure*}
\begin{center}
\includegraphics[width=1\textwidth]{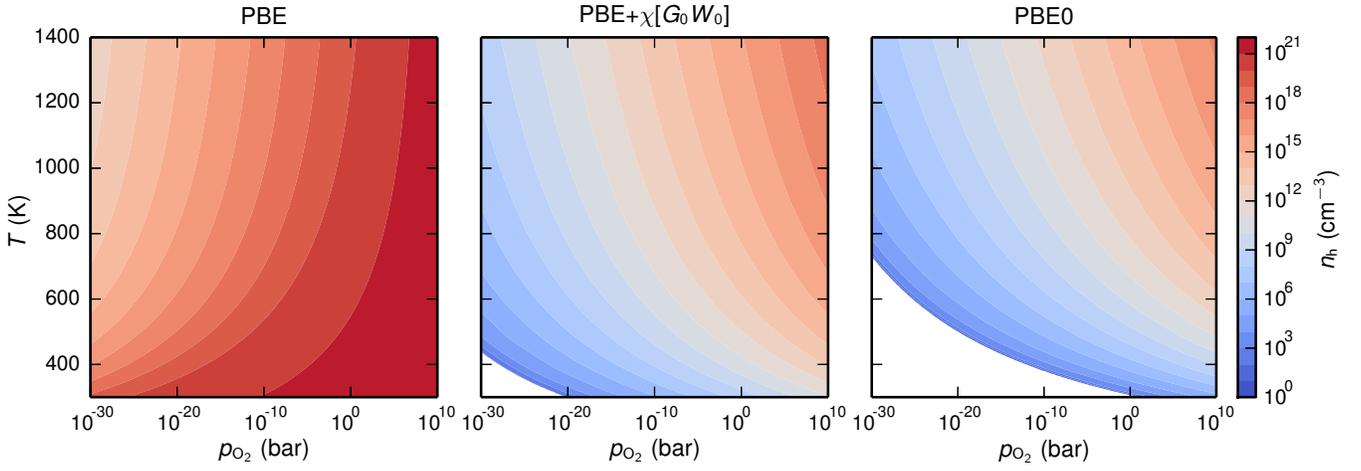}
\caption{Hole concentration calculated based on PBE, PBE0 and \PBExGW\ data under dry conditions at different temperatures and oxygen partial pressures. The dopant concentration is 10\%, which corresponds to $\unit[1.3\times10^{21}]{cm^{-3}}$.}
\label{fig:oxconc}
\end{center}
\end{figure*}

We have also studied the dry system for a wide range temperatures and oxygen partial pressures. 
In \fig{fig:oxconc} we show the hole concentration for different temperatures and oxygen partial pressures at a dopant concentration of 10\%. 
The holes completely compensate the acceptor dopants at high partial pressures if PBE energies are used, and the hole concentration is still quite substantial when the pressure decreases.
With \PBExGW\ and PBE0 we obtain a different picture. 
Here does the hole concentration become large only at high temperatures and very high partial pressures, and consequently the acceptor dopants are compensated by oxygen vacancies over most of the considered range.

\subsection{Hydration}
In the same manner as for the oxidation reaction, the hydration reaction can be studied through the corresponding equilibrium constant
\begin{equation}
\Khydr(T) = \left(\frac{p_{\HtO}}{p_{\HtO}^{\circ}}\right)^{-1}\frac{c_\H^2}{c_\V c_\O }, 
\end{equation}
where $c_\H$ is the proton concentration. 
In this case the equilibrium constant can be written as \cite{supplementary} 
\begin{equation}\label{eq:khydr1}
\Khydr(T) = \left(\frac{N_\H}{N_\V}\right)^2e^{-\Hhydr(T)/\kbt}e^{\Shydr(T)/\kb}.
\end{equation}
The difference in the number of sites available for protons ($N_\H$) and oxygen vacancies ($N_\V$) introduces an additional configurational contribution to the entropy \cite{bjorheim_hydration_2011,bjorheim_hydration_2015} and we can define an effective hydration entropy according to  
\begin{equation}
\Shydreff(T) = \Shydr(T) + \kb\ln{\left(\frac{N_\H}{N_\V}\right)^2}.
\end{equation}
In the present case we have $N_\H = 4N_\V$ and the additional term is equal to \unit[0.24]{meV/K}. 

At \unit[900]{K} we obtain hydration enthalpies of $\unit[-0.68]{eV}$ with PBE and \PBExGW, and $\unit[-0.76]{eV}$ with PBE0. 
The corresponding effective hydration entropy at the same temperature is $\unit[-1.38]{meV/K}$.

\subsection{Experimental conditions}
The environmental conditions in experimental studies are often such that both hydration and oxidation take place simultaneously. 
This is the case for a hydrated material under oxidizing conditions and during such circumstances it is not possible to consider the two reactions independently.

\begin{figure*}
  \centering
  \includegraphics[width=\textwidth]{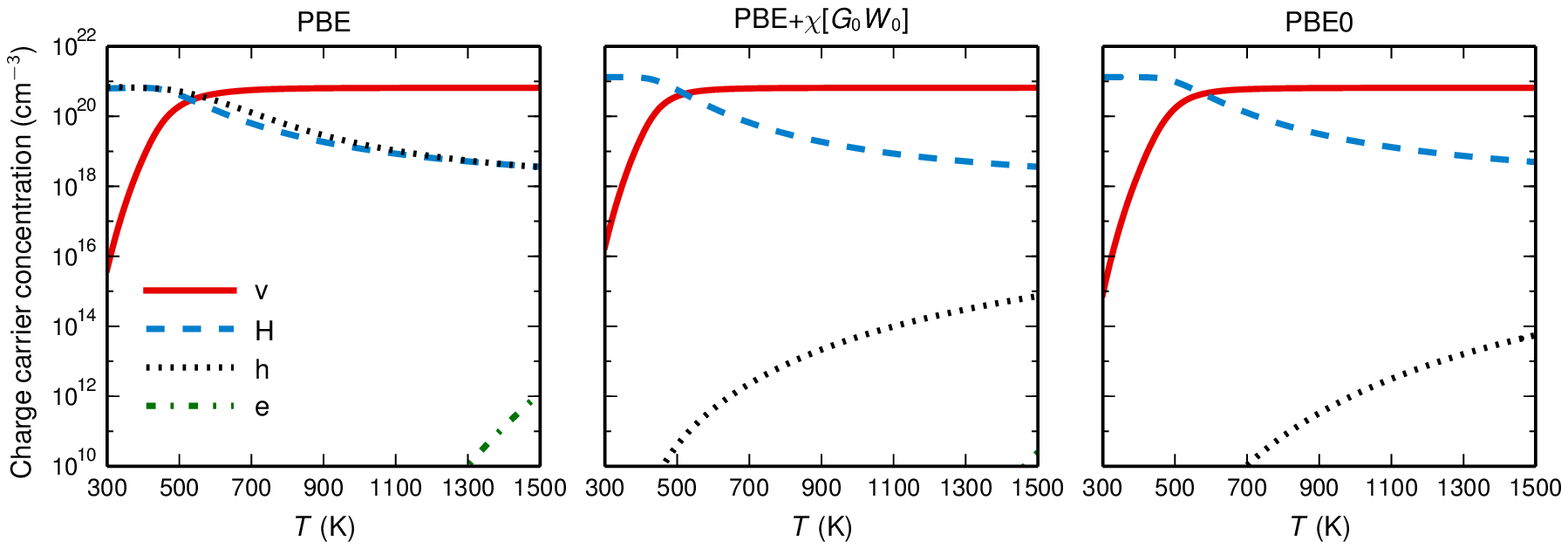}
  \caption{
    Concentration profiles calculated based on PBE, PBE0 and \PBExGW\ data under hydrated conditions at $p_{\HtO}=\unit[0.02]{bar}$ and $p_{\Ot}=\unit[10^{-5}]{bar}$. The dopant concentration is 10\%, which corresponds to $\unit[1.3\times10^{21}]{cm^{-3}}$.
}
  \label{fig:conc}
\end{figure*}

We have employed the scheme described in \Sect{sec:theory} to model these experimental conditions.
Concentration profiles for a 10\% doped material under wet conditions with $p_{\HtO}=\unit[0.02]{bar}$ and $p_{\Ot}=\unit[10^{-5}]{bar}$ are shown in \fig{fig:conc}. 
The material is hydrated at lower temperatures according to all three methods but only completely protonated for PBE0 and PBE+\GW. 
With PBE the hydration occurs in competition with hole formation leading to a situation with roughly 50\%\ protons and 50\%\ holes. 
Similar to dry conditions, the hole concentration increases with increasing temperature for both PBE0 and \PBExGW\ while the behavior is the opposite for PBE. 

In this study only isolated defects are considered, which is reasonable for low dopant concentrations. 
However, at higher concentrations defect ordering and association effects cannot be neglected. 
Real systems are often subject to high dopant concentrations of approximately 20\%\ and above.
While in such situations defect-defect interactions should be included we do not consider this complication in the present work.
The scheme employed here (\Sect{sec:theory}) can, however, be extended in straightforward fashion to account for additional defect species, including defect pairs, as a first order approximation to defect-defect interactions.

\section{Discussion}\label{sec:discussion}
\subsection{Hydration}
It was established in the previous section that the three methods considered in this work all predict very similar results for the hydration reaction. 
This shows that a change in description of the electronic structure has a small effect on the hydration enthalpy. 

The hydration of acceptor-doped \BZO\ has been studied extensively by several experimental groups and a compilation of their results is provided in \tab{tab:Hydrcomp}. 
There is good agreement between the different experimental results obtained at higher temperatures with some slight differences due to doping, where the dopant species appears to have a more prominent impact on the results compared to the dopant concentration. 

\begin{table*}
\centering
\caption{Experimental values of hydration enthalpies and entropies for various acceptor-doped \BZO\ systems.}
\begin{ruledtabular}
\begin{tabular}{l e{3.4} d{1.3} p{4.3} p{4.3} l }
\multicolumn{1}{c}{System} & \multicolumn{1}{c}{$T$ (K)} & \multicolumn{1}{c}{$p_{\HtO}$ (bar)} & \multicolumn{1}{c}{\Hhydr\ (eV)} & \multicolumn{1}{c}{\Shydr\ (meV\thinspace K$^{-1}$)} & \multicolumn{1}{c}{Reference}   \\ \hline
BaZr$_{0.98}$Y$_{0.02}$O$_{3-\delta}$ & 773-1173 & 0.023 & -0.84 & -0.98  & Kreuer \etal \cite{kreuer_proton_2001}  \\
BaZr$_{0.95}$Y$_{0.05}$O$_{3-\delta}$ & 773-1173 & 0.023 & -0.82 & -0.97  & Kreuer \etal \cite{kreuer_proton_2001}  \\
BaZr$_{0.9}$Y$_{0.1}$O$_{3-\delta}$ & 773-1173 & 0.023 & -0.82 & -0.92  & Kreuer \etal \cite{kreuer_proton_2001}  \\
 & 773-1073 &  \multicolumn{1}{c}{0.005-0.04} & -0.77,0.03 & -0.90,0.10  & Schober \& Bohn \cite{schober_water_2000}  \\ 
 & 573-1173 &  \multicolumn{1}{c}{0.1-1.0} & -0.84,0.04 & & Kj{\o}lseth \etal \cite{kjolseth_determination_2010}  \\  
 & 673-873 &  0.02 & -0.86 & -0.95 & Ricote \etal \cite{ricote_water_2009}  \\ 
BaZr$_{0.85}$Y$_{0.15}$O$_{3-\delta}$& 773-1173 &  0.023 & -0.86 & -0.95  & Kreuer \etal \cite{kreuer_proton_2001}  \\
BaZr$_{0.8}$Y$_{0.2}$O$_{3-\delta}$& 773-1173 & 0.023 & -0.97 & -1.07  & Kreuer \etal \cite{kreuer_proton_2001}  \\
 & 323-773 &  0.023 & -0.23,0.01 & -0.40,0.01 & Yamazaki \etal \cite{yamazaki_defect_2008}  \\ 
 & 773-1173 &  0.023 & -0.73 & -1.04 & Yamazaki \etal \cite{yamazaki_defect_2008}  \\ 
BaZr$_{0.9}$Sc$_{0.1}$O$_{3-\delta}$& 773-1173 &  0.023 & -1.24 & -1.29  & Kreuer \etal \cite{kreuer_proton_2001}  \\
BaZr$_{0.9}$Gd$_{0.1}$O$_{3-\delta}$& 773-1173 &  0.023 & -0.69 & -0.89  & Kreuer \etal \cite{kreuer_proton_2001}  \\
BaZr$_{0.9}$In$_{0.1}$O$_{3-\delta}$& 773-1173 &  0.023 & -0.69 & -0.93  & Kreuer \etal \cite{kreuer_proton_2001}  \\
\end{tabular}
\end{ruledtabular}
\label{tab:Hydrcomp}
\end{table*}

At $T=\unit[900]{K}$ the calculated hydration enthalpy is $\Hhydr=\unit[-0.68]{eV}$ with PBE and \PBExGW, and $\Hhydr=\unit[-0.76]{eV}$ with PBE0. 
Since these values are computed for an effectively acceptor-doped \BZO\ system there is no specific entry in \tab{tab:Hydrcomp} to compare with, although the values do agree quite well in general.
For the same temperature the calculated effective hydration entropy is $\Shydreff=\unit[-1.38]{meV/K}$. 
The magnitude of this value is somewhat larger than the experimental entropies listed in \tab{tab:Hydrcomp}. 
Recent investigations have shown that a more accurate treatment of the lattice vibrations gives a considerably better agreement with experiments \cite{bjorheim_hydration_2015}.

There is one entry in \tab{tab:Hydrcomp}, which differs from the others, namely the 20\%\ yttrium-doped system studied at low temperatures by Yamazaki \etal\cite{yamazaki_defect_2008}. 
The absolute value of the enthalpy is much smaller in this case, which corresponds to a less exothermic reaction. 
The authors argue that the difference with respect to other results is that the hole concentration can be neglected at low but not at high temperatures. 
This explanation is not consistent with either the PBE or the PBE0/\PBExGW\ results in \fig{fig:conc}.
Kj{\o}lseth \etal\ \cite{kjolseth_determination_2010} on the other hand argue that the less exothermic behavior is due to association and ordering between defects and dopants, under the assumption that oxygen vacancies are more associated and ordered compared to protons. 

\subsection{Oxidation}
The modeling of the oxidation reaction yields very different results depending of the method that is considered. 
The values of the oxidation enthalpy in \tab{tab:eformcomp} show that the standard DFT approach based on the PBE exchange-correlation functional predicts the reaction to be exothermic while \PBExGW\ and PBE0 predict an endothermic behavior. 

Figures~\ref{fig:oxconc} and \ref{fig:conc} show that the exothermic nature of the PBE results yields large hole concentrations. 
The results for PBE in the latter figure indicate that 50\%\ of the oxygen vacancies are oxidized even under hydrated conditions.
This is inconsistent with experiments, where almost completely hydrated samples are obtained \cite{kreuer_proton_2001,kreuer_proton_2003}. 

Unlike for the hydration reaction, there are to our knowledge no reported experimental values of the oxidation enthalpy for \BZO\ systems in the literature. 
There are however experimental values for other perovskite oxides, namely BaCeO$_3$\cite{he_impedance_1997}, BaTiO$_3$\cite{chan_nonstoichiometry_1981} and SrTiO$_3$\cite{fleischer_hole_1992}. 
The oxidation enthalpies for these systems (see \tab{tab:Hoxcomp}) are all positive, which corresponds to the reaction being endothermic. 

\begin{table*}
\begin{center}
\caption{Lattice constants $a_0$,  band gaps $\Egap$, band edge shifts and oxidation enthalpies \Eox/\Hox\ for several perovskite oxides. All calculations have been performed with the cubic perovskite structure. For the band edge shifts it is assumed that the PBE and PBE0 band structures are aligned. Energies and lattice constants are given in units of eV and \AA, respectively.}
\begin{ruledtabular}
\begin{tabular}{l c d{1.4} c c d{1.3} c c c c c }
			& \multicolumn{2}{c}{$a_0$}	& \multicolumn{3}{c}{$\Egap$}  & \multicolumn{2}{c}{Band-edge shifts}					& \multicolumn{2}{c}{\Eox}  & \Hox
\\  \cline{2-3}\cline{4-6}\cline{7-8}\cline{9-10}\cline{11-11}
 \multicolumn{1}{c}{System}		& PBE	& \multicolumn{1}{c}{Exp.}			& PBE	& PBE0	& \multicolumn{1}{c}{Exp.}			& VBM	& CBM	& PBE	& PBE0 	& Exp.									\\ \hline
BaCeO$_3$	& 4.476	& 4.445 \cite{knight_structural_1994}		& 2.25	& 4.95	& 4.41 \cite{he_optical_1996}		& $-1.45$ 			& $1.25$			& $-1.35$	& 1.85 & 1.11 \cite{he_impedance_1997}			\\
BaTiO$_3$	& 4.031	& 3.991 \cite{edwards_structure_1951} 	& 1.71 	& 3.82 	& 3.21 \cite{wemple_polarization_1970}		& $-1.41$			& $0.70$ 			& $-1.07$	& 1.96 & 0.92 \cite{chan_nonstoichiometry_1981} 	\\
SrTiO$_3$	& 3.939	& 3.900 \cite{abramov_chemical_1995} & 1.81	& 3.98	& 3.25 \cite{benthem_bulk_2001}		& $-1.41$			& $0.77$			& $-1.00$ & 1.87 & 1.40\cite{fleischer_hole_1992}			\\
\BZO			& 4.235	& 4.191\cite{levin_phase_2003}	& 3.13	& 5.35	& \multicolumn{1}{c}{4.8-5.3 \cite{robertson_band_2000,cavalcante_experimental_2008,cavalcante_intense_2008,cavalcante_intense_2009,yuan_synthesis_2008}	}	& $-1.42$			& $0.80$			& $-1.31$ & 1.53 &									\\
\end{tabular}
\end{ruledtabular}
\label{tab:Hoxcomp}
\end{center}
\end{table*}

To compare these experimental values with theoretical predictions \Eox\ was calculated for these perovskites as well. 
Calculations were performed with both PBE and PBE0 using the same computational setup as for \BZO. 
Band gaps and band edge shifts were determined as well, where the latter were obtained under the assumption that the PBE and PBE0 band structures are aligned. 
Although the cubic perovskite structure is not the ground state for these materials it was chosen for simplicity. 

The results of these calculations are shown in \tab{tab:Hoxcomp}. 
These three perovskites behave qualitatively similar to \BZO\ with negative and positive oxidation energies with PBE and PBE0, respectively. 
The latter are in better agreement with the experimental data.
The band gaps are also improved for all systems and the VBM and CBM are shifted downwards and upwards respectively for all materials, similar to \BZO. 
The fact that the overall improvement of PBE0 over PBE is a general feature for these three systems in combination with their similarities to \BZO\ strongly suggests an endothermic oxidation reaction in \BZO.
Thus, going beyond standard DFT is a necessary procedure when studying the oxidation reaction in these materials.
 
Throughout this article we have considered the hole to be a delocalized band state.
If the hole instead would be a localized polaronic state (small polaron) then the oxidation enthalpy would be reduced by the formation energy of the polaron.
Recent theoretical studies\cite{erhart_efficacy_2014,chen_hole_2014} based on the HSE functional and LDA$+U$ show indeed that polaron formation is favorable in several perovskites (SrTiO$_3$, BaTiO$_3$ and CaTiO$_3$).
However, the polaron formation energies are only about \unit[0.1-0.2]{eV} and thus quite small.
While polaron formation would reduce \Eox\ by \unit[0.2-0.4]{eV} it would not change the main conclusions of the paper.

\subsection{Conductivity}

Conductivity is a quantity that can be experimentally measured much more easily than defect concentrations. 
The conductivity of a charge carrier $i$ can be decomposed into 
\begin{equation}\label{eq:conductivity}
\sigma_i = q_iB_i n_i, 
\end{equation}
where $q_i$ is the carrier charge, $B_i$ the mobility and $n_i$ is the carrier concentration.  

\begin{figure}
\begin{center}
\includegraphics[width=0.48\textwidth]{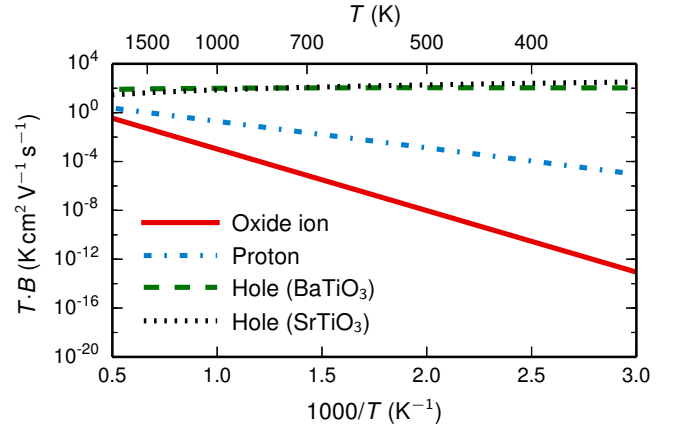}
\caption{
  Experimental mobility of charge carriers. The proton and oxide ion mobility is for BaZr$_{0.9}$Y$_{0.1}$O$_{3-\delta}$ and is based on data from Ref.~\onlinecite{kreuer_proton_2001}. The hole mobilities are based on the expressions given in Ref.~\onlinecite{chan_nonstoichiometry_1981} (BaTiO$_3$) and Ref.~\onlinecite{fleischer_hole_1992} (SrTiO$_3$).
}
\label{fig:mobility}
\end{center}
\end{figure}

Total and partial conductivities of yttrium-doped \BZO\ have been determined experimentally by several research groups\cite{kreuer_proton_2001,bohn_electrical_2000,wang_ionic_2005,nomura_transport_2007,kuzmin_total_2009}. 
With a fit to the Arrhenius like expression 
\begin{equation}\label{eq:arrhenius}
T\sigma_\h = Ae^{-E_a/\kbt}
\end{equation}
the reported hole conductivities $\sigma_\h$ yield activation energies $E_a$ in the range \unit[0.62]{eV} to \unit[1.05]{eV} \cite{bohn_electrical_2000,wang_ionic_2005,nomura_transport_2007,kuzmin_total_2009}. 
To compare the experimental hole conductivities with our results for the hole concentrations the mobility of the holes is required. 
While the mobility and hence the diffusion coefficient have been experimentally determined for both protons and oxygen vacancies in yttrium-doped \BZO, the hole mobility is unknown. 
There are, however, mobilities reported in the literature for other perovskites including BaTiO$_3$ \cite{chan_nonstoichiometry_1981} and SrTiO$_3$ \cite{fleischer_hole_1992}. 
In \fig{fig:mobility} these hole mobilities are depicted together with the proton and oxygen ion mobility in BaZr$_{0.9}$Y$_{0.1}$O$_{3-\delta}$ based on experimental data from Kreuer \etal\cite{kreuer_proton_2001}. 
Unlike the proton and oxygen ion mobilities, which clearly show temperature activated behavior, the hole mobilities have a temperature dependence close to $T^{-1}$ corresponding to scattering limited band conduction mechanism.

By assuming that $B_\h\sim T^{-1}$ it follows from \Eq{eq:conductivity} and \Eq{eq:arrhenius} that $n_\h \sim e^{-E_a/\kbt}$. 
If we consider \PBExGW\ and PBE0, where $n_\h \ll c_\V$ and the Boltzmann approximation is valid, we get \cite{supplementary}
\begin{equation}\label{eq:activationenergy}
E_a = \frac{\Hox(T) + 3\kbt}{2}.
\end{equation}
At $T=\unit[1000]{K}$ the calculated oxidation enthalpies yield $E_a = \unit[0.65]{eV}$ and $E_a = \unit[0.96]{eV}$ for \PBExGW\ and PBE0, respectively, which are within the range of the experimental results\cite{bohn_electrical_2000,wang_ionic_2005,nomura_transport_2007,kuzmin_total_2009}.

On the other hand, if we consider PBE the oxidation reaction is exothermic and $E_a$ is negative (c.f.~\fig{fig:kox}).
This can not be made consistent with the measured conductivity under the assumption of a weakly temperature dependent mobility, $B_\h\sim T^{-1}$.
For the PBE result to become consistent one has to assume a strongly temperature dependent mobility.
 In Refs.~\onlinecite{hermet_thermodynamics_2012,bevillon_oxygen_2011} it was suggested that the hole conductivity is given by a thermally activated process involving small polarons with a mobility given by $B_\h \sim T^{-1}\exp{\left( -E_{\text{mig}}/\kbt \right)}$. In the present case the activation energy for hole migration $E_{\text{mig}}$ has to be at least \unit[1]{eV}, which is unlikely.
 
\section{Summary and conclusions}\label{sec:summary}
In the present work we have studied the oxidation and hydration of an acceptor-doped proton-conducting perovskite oxide, \BZO, in contact with water vapor and oxygen gas. 
Charge carrier concentrations have been determined for different temperatures and partial pressures based on data from first-principles modeling.

Two different methods have been employed that improve upon the conventional PBE functional with regard to the description of band gap and band edges, namely the PBE0 hybrid functional and PBE+\GW\ calculations rooted in many-body perturbation theory.

We find that the hydration reaction is exothermic and well described by both PBE and PBE0. 
Including the band edge shifts from \GW\ calculations (\PBExGW) does not change the energetics for the hydration reaction.

For the oxidation reaction, however, the different approximations predict {\em qualitatively} different results. 
With PBE the reaction becomes exothermic while it is endothermic with PBE0 and \PBExGW. 
The exothermic PBE behavior yields large hole concentrations when lowering the temperature even under hydrated conditions and the oxide can not become completely hydrated, in disagreement with experiments. 
For the exothermic nature of PBE to be consistent with the experimental data for the hole conductivity the hole mobility has to increase more rapidly than the decrease in hole concentration. 
Such a temperature dependent hole mobility is unlikely.
We conclude that only the endothermic behavior with PBE0 and PBE+\GW\ can be made consistent with experimental data of charge carrier concentrations and hole conductivities.

In summary, PBE gives a good description for the hydration reaction but to model the oxidation reaction improved approximations have to be used. 
Here we show that the PBE+\GW\  method and hybrid functionals are two viable alternatives and we present a theoretical approach, which in a consistent way describes both hydration and oxidation of proton conducting acceptor-doped perovskites.

\begin{acknowledgments}
We would like to acknowledge the Swedish Energy Agency for financial support (Project number: 36645-1). Computational resources have been provided by the Swedish National Infrastructure for Computing (SNIC) at Chalmers Centre for Computational Science and Engineering (C3SE) and National Supercomputer Centre (NSC).
\end{acknowledgments}

\appendix*

\section{Band structure alignment}\label{sec:appendix}

\begin{figure}
\begin{center}
\includegraphics[width=0.48\textwidth]{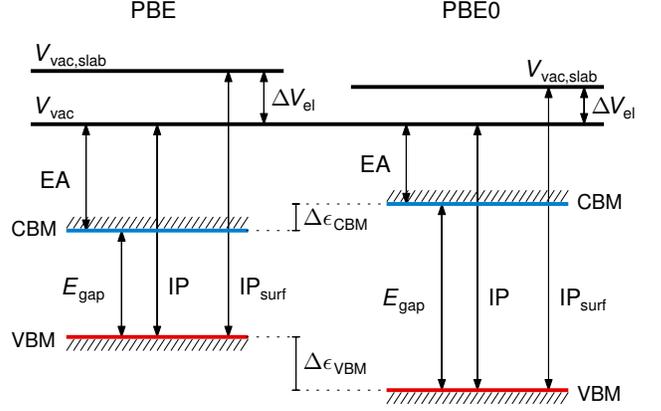}
\caption{Schematic representation of the band structure alignment between PBE and PBE0.}
\label{fig:vacuum_level}
\end{center}
\end{figure}

To determine the shift of the VBM and CBM between PBE and PBE0 the band structures need to be aligned. 
Such an alignment can be done with respect to a common reference potential, such as the vacuum level $V_{\text{vac}}$\cite{alkauskas_defect_2008,alkauskas_defect_2011}.
A schematic representation of the alignment is depicted in \fig{fig:vacuum_level}.
If the band structures are aligned with respect to this reference then the shifts of the VBM and CBM are given by the difference in the ionization potential IP and electron affinity EA, respectively, 
\begin{align}\label{eq:vbm_ip}
\Delta\Evbm &= \IP^{\text{PBE}} - \IP^{\text{PBE0}} \\
\label{eq:cbm_ea}
\Delta\Ecbm &= \EA^{\text{PBE}} - \EA^{\text{PBE0}},
\end{align}
where $\IP = V_{\text{vac}} - \Evbm$ and $\EA = V_{\text{vac}} - \Ecbm$.

To determine $V_{\text{vac}}$ and consequently IP and EA a surface calculation has to be performed.
Such a calculation requires a supercell containing a sufficiently long slab of \BZO\, so that the core of the slab becomes bulk-like, as well as enough of vacuum, in order to reach the vacuum level. 
An important aspect of this approach is that the vacuum level of the slab system, $V_\text{vac,slab}$, is not the same as the desired vacuum level due to ionic and electronic relaxation at the surface of the slab and thus can not directly be used as vacuum level in the alignment procedure.
To obtain the actual vacuum level these surface contributions need to be removed:
\begin{equation}
V_\text{vac} = V_\text{vac,slab} -\Delta V_\text{el} - \Delta V_\text{ion},
\end{equation}
where $\Delta V_\text{el}$ and $\Delta V_\text{ion}$ are contributions from electronic and ionic relaxation at the surface, respectively.
In the following only electronic relaxation is considered, hence $\Delta V_\text{ion}=0$.
The desired IP can thus be extracted from the slab system according to
\begin{equation}
\IP = V_\text{vac,slab} - \epsilon_\text{VBM,slab} -\Delta V_\text{el} = \IP_\text{surf} -\Delta V_\text{el}
\end{equation}
and together with \Eq{eq:vbm_ip} we obtain the shift of the VBM according to
\begin{equation}\label{eq:vbm_ip_calc}
\Delta \Evbm = \IP_\text{surf}^{\text{PBE}} - \IP_\text{surf}^{\text{PBE0}} - (\Delta V_\text{el}^{\text{PBE}}-\Delta V_\text{el}^{\text{PBE0}}).
\end{equation}
In the same manner we obtain the following expression for the CBM shift
\begin{equation}\label{eq:cbm_ea_calc}
\Delta \Ecbm = \EA_\text{surf}^{\text{PBE}} - \EA_\text{surf}^{\text{PBE0}} - (\Delta V_\text{el}^{\text{PBE}}-\Delta V_\text{el}^{\text{PBE0}}).
\end{equation}

\begin{figure}
\begin{center}
\includegraphics[width=0.48\textwidth]{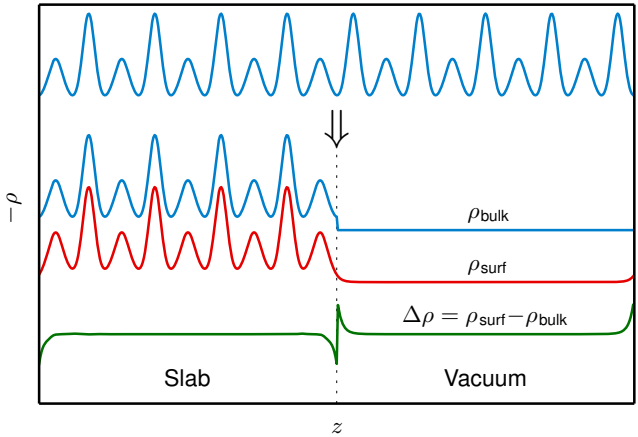}
\caption{Schematic representation of how the surface dipole charge density $\Delta\rho$ is obtained. The $\Delta\rho$-curve (green) has been multiplied with a factor of 5 for clarification.}
\label{fig:surface_dipole}
\end{center}
\end{figure}

The electronic relaxation at the surface gives rise to a surface dipole (see \fig{fig:surface_dipole}).
If we denote the difference in the planar averaged (in the $xy$-plane) charge density between the surface and bulk systems $\Delta\rho(z)$, where $z$ is the axis perpendicular to the surface, then the potential arising from the surface dipole can be calculated from the expression \cite{heimel_interface_2008}
\begin{equation}
\Delta V_\text{el} = -\frac{p}{\varepsilon_0 A},
\end{equation}
where $\varepsilon_0$ is the vacuum permittivity, $A$ is the unit area, and $p$ is the electric dipole moment  
\begin{equation}
p = \int \Delta\rho(z)(z-z_0)dz,
\end{equation}
with $z_0$ denoting the center of mass.

In an actual calculation $\Delta\rho(z)$ is obtained in the following manner (for a schematic representation see \fig{fig:surface_dipole}). 
First, a $1\times1\times n$ supercell is constructed and the corresponding charge density $\rho_\text{bulk}(z)$ is determined.
Half of the atoms are then removed resulting in a supercell containing a $1\times1\times \tfrac{n}{2}$ slab and an equal amount of vacuum.
Subsequent electronic relaxation yields the charge density $\rho_\text{surf}(z)$.
$\Delta\rho(z)$ is then obtained as the difference between these charge densities,
\begin{equation}
\Delta\rho(z) = \rho_\text{surf}(z) - \rho_\text{bulk}(z),
\end{equation}
where $\rho_\text{bulk}(z)$ has been truncated and set to zero at the same position as the surface in the slab supercell.

To determine the VBM and CBM shifts we have considered the $[001]$ surface with both ZrO$_2$ and BaO terminations.
We have used $n=9$, which corresponds to a slab consisting of four and a half unit cells, where both surfaces (the second surface arise from the periodic boundary conditions) have the same termination. 
We use the same computational setup as described in \Sect{sec:comp}, however, only one $k$-point is used in the $z$-direction. 
A summary of the results is given in \tab{tab:alignment}.
Using \Eq{eq:vbm_ip_calc} we obtain VBM shifts of $\unit[-1.44]{eV}$ and $\unit[-1.38]{eV}$ for the ZrO$_2$ and BaO-terminated surfaces, respectively.
These shifts are in very good agreement with the VBM shift of $\unit[-1.42]{eV}$ obtained by directly comparing PBE and PBE0 results for the bulk.
For the CBM shifts we obtain $\unit[0.83]{eV}$ for the ZrO$_2$-terminated surface, which compares well with the direct value of \unit[0.80]{eV}.
For the BaO-terminated surface, however, the CBM shift is only \unit[0.34]{eV}.
This discrepancy is likely related to the fact that the conduction band consists of zirconium $d$-states (see \fig{fig:pdos_comp}), which are not present in the surface layer for the BaO termination. 
In all, the results obtained here demonstrate the proper alignment of PBE and PBE0 band structures (at identical lattice constant and using the same pseudopotentials) shown in \fig{fig:bs_ideal_comp}.

\begin{table}[ht!]
\begin{center}
\caption{Difference between PBE and PBE0 results for the bulk system as well as both terminations of the [001] surface. Equivalent band edge shifts for the different systems are given in bold. Energies are given in units of eV.}
\begin{ruledtabular}
\begin{tabular}{l d{2.2} d{2.2} d{2.2} }
Quantity		& \multicolumn{1}{c}{Bulk}	& \multicolumn{1}{c}{ZrO$_2$}	& \multicolumn{1}{c}{BaO} 	\\ \hline
$\Delta\Evbm$	& \bf -1\bf.\bf4\bf2	& -1.37		& -1.26	\\
$\Delta\Ecbm$ & \bf0\bf.\bf8\bf0	& 0.90		& 0.36	\\
$\Delta\Egap$	& 2.22	& 2.26		& 1.62	\\
$ V_\text{vac}^{\text{PBE}}-V_\text{vac}^{\text{PBE0}}$	& 	& 0.1	0	& 0.14	\\
$-(\Delta V_\text{el}^{\text{PBE}}-\Delta V_\text{el}^{\text{PBE0}}$)	& 	& -0.17		& -0.27	\\
$\Delta\Evbm$ from \Eq{eq:vbm_ip_calc}	& 	& -\bf1\bf.\bf4\bf4		& -\bf1\bf.\bf3\bf8	\\
$\Delta\Ecbm$ from \Eq{eq:cbm_ea_calc}	& 	& \bf0\bf.\bf8\bf3		& \bf0\bf.\bf3\bf4	\\
\end{tabular}
\end{ruledtabular}
\label{tab:alignment}
\end{center}
\end{table}

\bibliography{oxidation_paper}

\end{document}


\title{Supplemental Material: Implications of the band gap problem on oxidation and hydration in acceptor-doped barium zirconate}

\author{Anders Lindman}
\author{Paul Erhart}
\author{G\"oran Wahnstr\"om}

\affiliation{
  Department of Applied Physics,
  Chalmers University of Technology,
  SE-412 96 Gothenburg, Sweden
}

\begin{abstract}
In this supplemental material we derive explicit expressions for the equilibrium constants of the hydration and oxidation reactions in acceptor-doped perovskite oxide materials.
These expressions are used in the main article to make connections between the theoretical modeling and experiments.
For the oxidation reaction we consider two limiting cases: \textit{high} and \textit{low} hole concentrations.
\end{abstract}

\maketitle
   
\section{Oxidation reaction}
Consider first the oxidation reaction
\begin{equation}\label{eq:oxidation}
\frac{1}{2}\Ot(\text{g}) + \VOpp \rightleftharpoons 2\h^{\bullet} + \Ox.
\end{equation}
This reaction is associated with the Gibbs free energy of formation
\begin{equation}\label{eq:gox}
\Gox(T,p_\Ot) = 2\mue(T)-\DGV,
\end{equation}
where $\mue(T)$ is the electron chemical potential with respect to the valance band edge and $\DGV$ is the Gibbs free energy of formation for a doubly charged oxygen vacancy.
The corresponding equilibrium constant is given by 
 \begin{equation}\label{eq:kox1}
\Kox(T) = \left( \frac{p_\Ot}{p_\Ot^{\circ}}\right)^{-1/2}\frac{n_\h^2c_\O}{c_\V},
\end{equation}
where $n_\h$, $c_\V$ and $c_\O$ denote hole, oxygen vacancy and oxygen ion concentrations, respectively, and $p_\Ot$ ($p_\Ot^{\circ}$) is the (reference) oxygen partial pressure. 

In the dilute limit the concentration of an arbitrary defect is given by
\begin{equation}\label{eq:conc}
c_{\text{def}} = \frac{N_{\text{def}}}{V_c}e^{-\Delta G_{\text{def}}/\kbt},
\end{equation}
where $N_\text{def}$ is the number of defect sites in the unit cell with volume $V_c$ and $\Delta G_{\text{def}}$ is the Gibbs free energy of defect formation.
For the oxygen vacancy concentration we get
\begin{equation}\label{eq:vac_conc}
c_\V = \frac{N_\V}{V_c}e^{\Goxo(T)/\kbt}e^{-2\mue(T)/\kbt}\left( \frac{p_\Ot}{p_\Ot^{\circ}}\right)^{-1/2},
\end{equation}
where 
\begin{equation}
\Goxo(T) = \Hox(T) -T\Sox(T)
\end{equation}
is the standard Gibbs free energy of formation for the oxidation reaction. 
The site restriction on the oxygen sublattice implies that 
\begin{equation}
c_\O + c_\V = \frac{N_\V}{V_c}
\end{equation}
and consistent with the dilute limit we have $c_\O = N_\V/V_c$, hence
 \begin{equation}\label{eq:kox2}
\Kox(T) = n_\h^2e^{2\mue(T)/\kbt}e^{-\Goxo(T)/\kbt}.
\end{equation}

The electron chemical potential is determined from the charge neutrality condition, which in general can be written as 
\begin{equation}
\sum_{\text{def}}q_{\text{def}}c_{\text{def}}(\mue) - n_\e(\mue) + n_\h(\mue) = 0,
\label{eq:charge_neutrality}
\end{equation}
where $n_\e$ is the electron concentration. 
In this case we have
\begin{equation}
2c_\V + n_\h = c_\A,
\end{equation}
where $c_\A = x_\A/V_c$ is the acceptor dopant concentration  and $x_\A$ is the fractional acceptor dopant concentration.

To obtain an explicit expressions for the equilibrium constant a model for the density of states (DOS) $g(\epsilon)$ has to be introduced.
We assume a simple parabolic expression for the valence band
\begin{equation}
g(\epsilon) = 
\begin{cases}
   g_0\sqrt{\Evbm - \epsilon},&  \epsilon <  \Evbm\\
    0,              & \epsilon > \Evbm
\end{cases}
\end{equation}
with
\begin{equation}
g_0 = \frac{1}{2\pi}\left( \frac{2\meff}{\hbar^2} \right)^{3/2},
\end{equation}
where $\meff$ is an effective mass for the hole. 

We will now consider two limiting situations; \textit{low} and \textit{high} hole concentrations. 

\subsection{Low hole concentration}
We first consider the situation where the hole concentration is low: $n_\h \ll n_\V$.
This is the situation when the \PBExGW\ and PBE0 approximations are used.
In this case the electron chemical potential will be located in the band gap, $\mue> 0$, and we will assume that $\mue \gg \kbt $. 
In that case the hole concentration can be obtained using the Boltzmann approximation
\begin{align}
n_\h (\mue) &= \int_{-\infty}^{\Evbm}g_0 \sqrt{\Evbm - \epsilon} \left[ 1-f(\epsilon,\mue) \right]d\epsilon  \nonumber \\
&= n_\text{VB}(T)e^{-\mue(T)/\kbt }, 
\end{align}
where $n_\text{VB}(T)$ is the effective state density for the valence band, which is given by
\begin{equation}
n_\text{VB}(T) = 2\left( \frac{\meff \kbt }{2\pi\hbar^2} \right)^{3/2}.
\end{equation}

With $n_\h \ll c_\V$ the charge neutrality condition implies that $2c_\V = c_\A$.
With \Eq{eq:vac_conc} we then get the following expression for the electron chemical potential
\begin{equation}
e^{2\mue(T)/\kbt } = \left( \frac{p_\Ot}{p_\Ot^{\circ}}\right)^{-1/2}\frac{2N_\V}{x_\A}e^{\Goxo(T)/\kbt }
\end{equation}
and the hole concentration is thereby given by  
\begin{equation}
n_\h(T,p_\Ot) = \left( \frac{p_\Ot}{p_\Ot^{\circ}}\right)^{1/4}\sqrt{\frac{x_\A}{2N_\V}} n_\text{VB}(T) e^{-\Goxo(T)/2\kbt }.
\end{equation}
Finally, for the equilibrium constant we obtain
 \begin{equation}\label{eq:kox3}
\Kox(T) = [ n_\text{VB}(T)]^2 e^{-\Hox(T)/\kbt}e^{\Sox(T)/\kb}.
\end{equation}

If we define an effective oxidation enthalpy
\begin{equation}
\Hoxeff(T) = -\kb\frac{d\ln{\Kox(T)}}{d(1/T)}
\label{eq:Hoexff}
\end{equation} 
we get from \Eq{eq:kox3} that
\begin{equation}\label{eq:dlnkox_low}
\Hoxeff(T) = \Hox(T) + 3\kbt,
\end{equation} 
where we have made use of the thermodynamic relation
\begin{equation}
\frac{d}{dT}\Hox (T) = T\frac{d}{dT}\Sox (T).
\end{equation}
In the same way, if we write 
\begin{equation}
n_\h = Ae^{-\Delta E_\h/\kbt}
\end{equation}
for the hole concentration, we get an effective energy 
\begin{equation}
\Delta E_\h^{\text{eff}}(T) = -\kb\frac{d\ln{n_\h(T)}}{d(1/T)} = \frac{\Hox(T) + 3\kbt}{2}.
\end{equation} 

\subsection{High hole concentration}
Now we consider the case with high hole concentration: $n_\h \gg c_\V$.
This is the situation using the PBE approximation at low temperatures. 
In this case the electron chemical potential will be located in the valence band, $\mue < 0$.
We define $\nue = -\mue$ and assume that $\nue \gg \kbt $.
In this case the hole concentration can be obtained using the Sommerfeld expansion
\begin{align}
n_\h (\nue) &= \int_{-\infty}^{\Evbm}g_0 \sqrt{\Evbm - \epsilon} \left[ 1-f(\epsilon,\nue) \right]d\epsilon \nonumber \\
&= g_0\frac{2}{3}\nue^{3/2} \left[1 + \frac{\pi^2}{8}\left( \frac{\kbt }{\nue} \right)^2 + \ldots \right].
\label{eq:holesommerfeld}
\end{align}

With $n_\h \gg c_\V$ the charge neutrality condition implies $n_\h = c_\A$ and using \Eq{eq:holesommerfeld} we get after some algebra the following expression for the electron chemical potential
\begin{equation}
\nue(T) = \alpha \left[1 - \frac{\pi^2}{12}\left( \frac{\kbt }{\alpha} \right)^2 \right]
\end{equation}
with $\alpha = \left( 3c_\A / 2g_0 \right)^{2/3}$. 
This further implies that
 \begin{equation}
\Kox(T) = c_\A^2e^{-\left[\Goxo(T) + 2\nue(T) \right]/\kbt}
\end{equation}
and 
\begin{equation}\label{eq:dlnkox_high}
\Hoxeff(T) = \Hox(T) + 2\alpha \left[1 + \frac{\pi^2}{12}\left( \frac{\kbt }{\alpha} \right)^2 \right].
\end{equation} 
The hole concentration gives rise to an effective oxidation enthalpy that is larger than the standard oxidation enthalpy $\Hox(T)$.

\section{Hydration reaction}
Next we consider the hydration reaction
\begin{equation}\label{eq:hydration}  
\HtO(\text{g}) + \VOpp + \Ox \rightleftharpoons 2\OHp.
\end{equation}
The Gibbs free energy of formation for this reaction is given by
\begin{equation}\label{eq:ghydr}
\Ghydr(T,p_\HtO) = 2\DGH-\DGV
\end{equation}
where $\DGH$ is the Gibbs free energy of formation for a proton and $p_\HtO$ is the water partial pressure.
The equilibrium constant of \Eq{eq:hydration} is given by
\begin{equation}
\Khydr(T) = \left(\frac{p_{\HtO}}{p_{\HtO}^{\circ}}\right)^{-1}\frac{c_\H^2}{c_\V c_\O }, 
\end{equation}
where $c_\H$ is the proton concentration.

The site restriction now also includes the proton concentration
\begin{equation}
c_\O + c_\V + c_\H = \frac{N_\V}{V_c},
\end{equation}
however, in the dilute limit we obtain the same expression as for the oxidation reaction, $c_\O = N_\V/V_c$.
Using \Eq{eq:conc} for $c_\V$ and $c_\H$ we obtain 
\begin{align}
\Khydr(T) &= \left(\frac{p_{\HtO}}{p_{\HtO}^{\circ}}\right)^{-1}\frac{N_\H^2}{N_\V^2}e^{-2\DGH/\kbt} e^{\DGV/\kbt} \nonumber \\
&= \left( \frac{N_\H}{N_\V} \right)^2e^{-\Ghydro(T)/\kbt},
\end{align}
where
\begin{equation}
\Ghydro(T) = \Hhydr(T) -T\Shydr(T)
\end{equation}
is the standard Gibbs free energy of formation for the hydration reaction.